\documentclass[prb,reprint, amsmath,amssymb,showkeys,superscriptaddress,floatfix]{revtex4-2}
\usepackage[breaklinks=true,colorlinks,citecolor=blue,linkcolor=blue,urlcolor=blue]{hyperref}
\usepackage[table,xcdraw]{xcolor}

\usepackage{epsfig,mathrsfs,color,latexsym,subfigure,marginnote,gensymb,}
\usepackage{graphicx}
\usepackage{xcolor}
\usepackage{booktabs}
\usepackage{appendix}
\usepackage{enumitem}
\usepackage{url}
\renewcommand{\BibitemShut}[1]{}

\begin{document}

\title{Deep Learning-Driven Prediction of Microstructure Evolution via Latent Space Interpolation}

\thanks{Code available on \href{https://github.com/Sachin-G13/Deep-Learning-Driven-Prediction-of-Microstructure-Evolution-via-Latent-Space-Interpolation/tree/main}{GitHub}.}
 \author{Sachin Gaikwad}
 \thanks{These authors contributed equally to this work.}
	\affiliation{School of Materials Science and Technology, Indian Institute of Technology (BHU), Varanasi, India}
\author{Thejas Kasilingam}
\thanks{These authors contributed equally to this work.}
	\affiliation{Department of Materials Science $\&$ Engineering, Indian Institute of Technology, Kanpur, Kanpur 208016, India}
\author{Owais Ahmad}
	\affiliation{Department of Materials Science $\&$ Engineering, Indian Institute of Technology, Kanpur, Kanpur 208016, India}
\author{Rajdip Mukherjee}
	\affiliation{Department of Materials Science $\&$ Engineering, Indian Institute of Technology, Kanpur, Kanpur 208016, India}
\author{Somnath Bhowmick}
\email{bsomnath@iitk.ac.in}
	\affiliation{Department of Materials Science $\&$ Engineering, Indian Institute of Technology, Kanpur, Kanpur 208016, India}
\date{\today}

\begin{abstract}
Phase-field models accurately simulate microstructure evolution, but their dependence on solving complex differential equations makes them computationally expensive. This work achieves a significant acceleration via a novel deep learning-based framework, utilizing a Conditional Variational Autoencoder (CVAE) coupled with Cubic Spline Interpolation and Spherical Linear Interpolation (SLERP). We demonstrate the method for binary spinodal decomposition by predicting microstructure evolution for intermediate alloy compositions from a limited set of training compositions. First, using microstructures from phase-field simulations of binary spinodal decomposition, we train the CVAE, which learns compact latent representations that encode essential morphological features. Next, we use cubic spline interpolation in the latent space to predict microstructures for any unknown composition. Finally, SLERP ensures smooth morphological evolution with time that closely resembles coarsening. The predicted microstructures exhibit high visual and statistical similarity to phase-field simulations. This framework offers a scalable and efficient surrogate model for microstructure evolution, enabling accelerated materials design and composition optimization.
\end{abstract}
\maketitle
\section{Introduction}

Artificial intelligence (AI) and machine learning (ML) have rapidly emerged as a transformative force across scientific disciplines, owing to the availability of high-performance computing and vast datasets generated over the past centuries. In the domain of materials science and engineering, AI and ML have demonstrated their potential over traditional methods by enabling automated analysis and prediction across scales~\cite{XIONG2020109203, Jennings2019}. Some prominent examples are interpreting and denoising microstructural images~\cite{DECOST201730, Holm2020, AHMAD2025114963}, linking atomic-scale imaging with material properties~\cite{Han2022}, materials discovery and manufacturing~\cite{PAPADIMITRIOU2024112793}, and predicting early signs of material failure~\cite{LI2021109726}.

The focus of the current research is microstructure modeling. Among the physics-based computational approaches, the phase-field method stands out as a robust framework to simulate complex microstructural phenomena such as solidification~\cite{HOTZER2015194, ZHAO20191044}, precipitate growth~\cite{MUKHERJEE20093947}, grain growth~\cite{PhysRevLett.86.842, CHANG201767}, and spinodal decomposition~\cite{CHAFLE2019236, SEOL20035173}. These simulations are governed by partial differential equations that evolve spatially and temporally, requiring fine discretization and substantial computational resources. Efforts to reduce the computational overhead have led to the integration of numerical optimizations and parallel processing techniques~\cite{Vondrousdoi:10.1177/1094342013490972, Miyoshi2017}.

Recent advances in deep learning have inspired a new class of data-driven surrogate models capable of approximating phase-field simulations with significantly reduced runtimes. These works mainly rely on generating the training data from phase-field simulations, compressing them to a low-dimensional latent space, followed by time series prediction models like recurrent neural networks (RNN), including the long short-term memory (LSTM) unit and the gated recurrent unit (GRU) to learn the microstructure evolution in latent space~\cite{MontesdeOcaZapiain2021, HU2022115128, ahmad2023accelerating, ahmad2024, TIWARI2025113518}. While these machine learning methods have shown great success in terms of accelerating the study of microstructure evolution, one needs to provide the models with some initial microstructures from the phase-field simulations to generate the subsequent microstructures. Moreover, due to error compounding, very long-term predictions are not possible using time series prediction models. 

Generative models are also very popular in the field of microstructure modeling, particularly for learning microstructure features, generating synthetic microstructures, and predicting time evolution. While generative models are known for speed, efficiency, diversity, and statistical accuracy, they are more data and resource-intensive. Models like generative adversarial network (GAN)~\cite{HARIBABU2023112512, ahmad2025arxiv} and denoising diffusion probabilistic models (DDPM)~\cite{AZQADAN2023119406, BENTAMOU2025113596} have shown excellent potential in the field of microstructure modeling.

\begin{figure*}
\includegraphics[width=\textwidth]{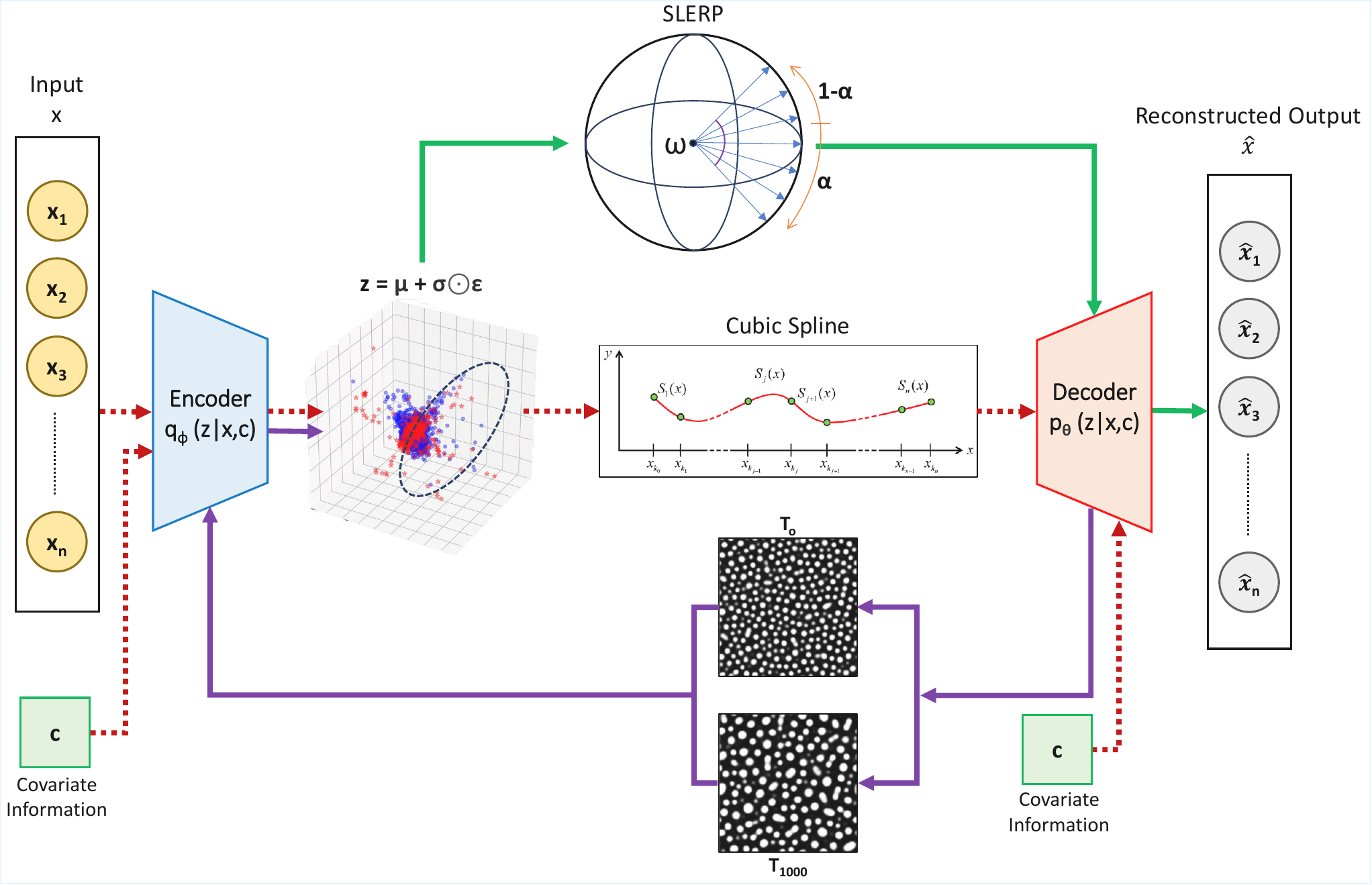}
\caption{Workflow of the Conditional Variational Autoencoder (CVAE) framework for modeling microstructure evolution. The CVAE takes as input a set of microstructure images \(x_1, x_2, ..., x_n\) along with their corresponding covariate information (c\textsubscript{avg}). The encoder maps the inputs to a latent space \(z = \mu + \sigma \odot \varepsilon\). The dotted red arrows represent the cubic spline interpolation workflow in the latent space, where representations for the targeted composition value are generated, which are then decoded to produce synthetic microstructures. The purple arrows show the selection process of latent vectors of images with the smallest and largest feature size (white shapes) and their injection back into the encoder. The green arrows depict the SLERP (Spherical Linear Interpolation) workflow, which operates between latent vectors corresponding to the microstructures with the largest and smallest feature size (white shapes) to generate a sequence of microstructures showing gradual and realistic morphological transitions of the targeted composition value.}
\label{fig:workflow}
\end{figure*}


This work uses Conditional Variational Autoencoder (CVAE)~\cite{kingma2013auto, SohnNIPS2015_8d55a249}, a generative model which learns a latent representation of input microstructures, conditioned on composition (c\textsubscript{avg}) values. To demonstrate our method, we deal with binary spinodal decomposition, a phase separation phenomenon that spontaneously separates a homogeneous mixture into two coexisting phases with different compositions. The initial average compositions lie within the interval of c\textsubscript{avg} = 0.27 to 0.48. The training set consists of 700 microstructures, each from nine selected compositions, 6300 images in total. Each grayscale microstructure has a dimension of $256\times256\times 1$. Given the high dimensionality of the dataset, the computational burden becomes significant in terms of both memory and processing time. The encoder compresses the data while preserving its essential features~\cite{Hintondoi:10.1126/science.1127647, WANG2016232} for further processing, which includes concatenating the input image with a learned label embedding, extracting features via convolutional layers, and mapping them to a lower-dimensional latent space with mean and log variance outputs. The reparameterization trick~\cite{kingma2013auto} is applied to generate latent vectors. The decoder reconstructs images by combining these vectors with class embeddings and upsampling via transposed convolution layers~\cite{Tang_2019_CVPR}. The loss function includes Mean Squared Error (MSE) for reconstruction and a KL divergence loss~\cite{kingma2013auto}, enforcing a structured latent space.

The workflow is illustrated in Figure~\ref{fig:workflow}. After compressing the data with the encoder, we use two types of latent space interpolations. The dotted red arrows represent the cubic spline interpolation, which generates representations for the targeted composition value. The decoder produces synthetic microstructures; the one with the smallest and largest feature size is chosen and is injected back into the encoder (purple arrows). Finally, spherical linear interpolation (green arrows) operates between latent vectors corresponding to the microstructures with the largest and smallest feature size, generating a smooth and realistic morphological evolution for a given targeted composition value. The rest of the paper provides technical details of various components of the workflow, followed by a discussion and a conclusion section.   

\section{Phase-Field method}
\label{sec:phase_field}
To generate a dataset for training and validation, we simulate spinodal decomposition in an A-B binary alloy using a phase-field framework, as implemented in the $\mu$2mech package~\cite{Linda_2024}. The evolution of microstructure is driven by minimizing the total free energy~\cite{Cahn10.1063/1.1744102}:
\[
F = \int_V \left[ f(c) + \kappa (\nabla c)^2 \right] dV,
\]
where \(c(\mathbf{r}, t)\) represents the conserved composition field, \(\kappa\) is the gradient energy coefficient. \(f(c)\) is the bulk free energy density modeled as a double-well potential,
\[
f(c) = W c^2 (1 - c)^2,
\]
where $W$ determines the height of the potential barrier. The free energy landscape favors phase separation when \(\frac{\partial^2 f}{\partial c^2} < 0\), indicating thermodynamic instability. The time evolution of the composition field follows the Cahn-Hilliard equation:
\[
\frac{\partial c}{\partial t} = M \left[ \nabla^2 \left( \frac{\partial f}{\partial c} \right) - 2\kappa \nabla^4 c \right].
\]
In the above equation, \(M\) is the mobility, taken as constant.

Nine different compositions are sampled, with initial average compositions of c\textsubscript{avg} = 0.27, 0.31, 0.35, 0.39, 0.40, 0.42, 0.44, 0.46, 0.48. All simulations are run on a 2D square domain with a grid size of \(256 \times 256\). Each simulation outputs a time series of microstructure images, evolved for 1000 time steps. However, due to the absence of significant features in the early evolution stages (\(t_0\) to \(t_{300}\)), we discard these frames and use data from \(t_{301}\) onward, capturing the critical period of phase separation and domain coarsening. Thus, 700 phase-field microstructures are obtained per composition, totalling 6300 images. This dataset is used to train our machine learning model, as detailed in the following sections. 

\section{Conditional Variational Autoencoder (CVAE) Architecture}
\label{sec:cvae}
\begin{figure}
    \centering
    \includegraphics[width=0.55\textwidth]{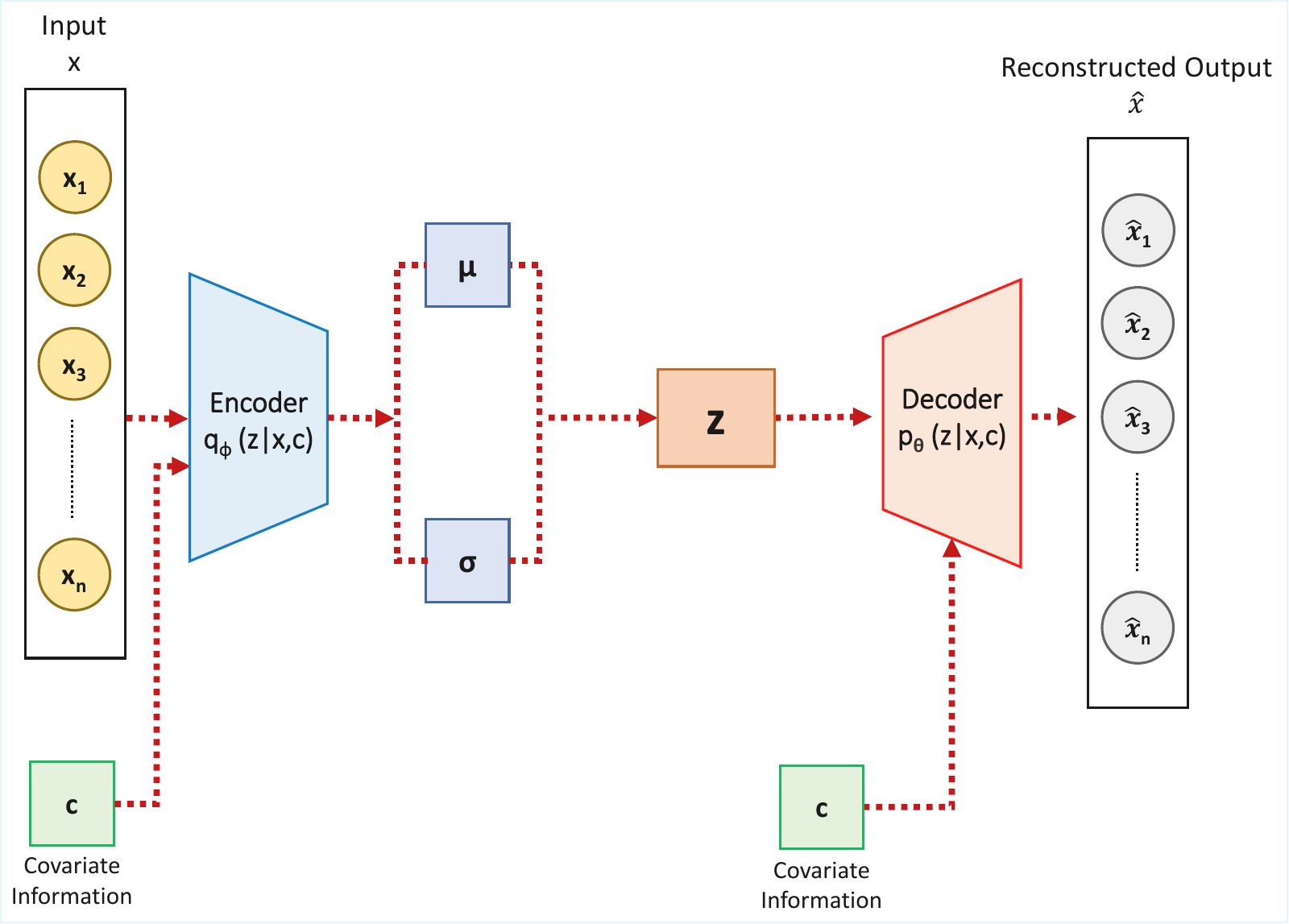}
    \caption{Schematic representation of the Conditional Variational Autoencoder (CVAE) architecture. The encoder maps the input \( x \) and covariate information \( c \) into a latent distribution parameterized by mean \( \mu \) and standard deviation \( \sigma \). A latent variable \( z \) is sampled and decoded back into the reconstructed output \( \hat{x} \) while conditioning on \( c \).}
    \label{fig:cvae_architecture}
\end{figure}

First, we need to map each image of size \(256 \times 256 \times 1\) to a lower-dimensional latent space that retains essential structural features while significantly reducing the data complexity. Dimensionality reduction techniques are widely used for this purpose, aiming to extract a compact set of features that retain the essential characteristics of the original data~\cite{Hintondoi:10.1126/science.1127647, WANG2016232}. In this study, we adopt a Conditional Variational Autoencoder (CVAE)~\cite{kingma2013auto, SohnNIPS2015_8d55a249}, a generative model that learns a latent representation of the input microstructure images while being conditioned on additional variables, specifically the average concentration (c\textsubscript{avg}) values. The CVAE consists of an encoder-decoder framework, wherein the encoder maps the input and the conditioning information to a structured latent space, and the decoder reconstructs the image from the sampled latent vectors and class embeddings. An overview of the CVAE architecture is illustrated in Figure~\ref{fig:cvae_architecture}.

\paragraph{Encoder Network:}
The encoder takes an input image $x \in \mathbb{R}^{H \times W \times C}$ along with its corresponding condition label $y \in \mathbb{R}^{N_c}$, where $H$, $W$, and $C$ represent the height, width, and number of channels of the image, respectively, and $N_c$ is the number of conditioning labels. The label is first embedded and reshaped to match the image dimensions,
\begin{equation}
y' = \text{Reshape}(\text{Dense}(H \times W)(y)).
\end{equation}
Then, the input image and the label embedding are concatenated:
\begin{equation}
z_{\text{input}} = \text{Concatenate}(x, y').
\end{equation}
Feature extraction is performed using a series of convolutional layers followed by max-pooling operations, resulting in a feature representation $z_{\text{conv}}$. The encoder then learns a Gaussian distribution in the latent space by computing the mean $\mu$ and log variance $\log \sigma^2$:
\begin{eqnarray}
\mu = \text{Dense}\left(\prod_{i=1}^{d} L_i\right)\left(z_{\text{conv}}\right),\\  
\log \sigma^2 = \text{Dense}\left(\prod_{i=1}^{d} L_i\right)\left(z_{\text{conv}}\right),
\end{eqnarray}
where $L_i$ represents the dimensions of the latent space.

\paragraph{Latent Space Sampling:}
To allow backpropagation, the reparameterization trick is applied, where a random variable $\epsilon \sim \mathcal{N}(0, I)$ is introduced:
\begin{equation}
z = \mu + \exp\left(\frac{\log \sigma^2}{2}\right) \cdot \epsilon.
\end{equation}

\paragraph{Decoder Network:}
The decoder reconstructs the original image given the latent variable $z$ and the conditional label $y$. The label is embedded similarly to the encoder:
\begin{equation}
y'' = \text{Reshape}\left(\text{Dense}\left(\prod_{i=1}^{d} L_i\right)(y)\right).
\end{equation}
The decoder input is then formed by concatenating $z$ with $y''$:
\begin{equation}
z_{\text{dec}} = \text{Concatenate}(z, y'').
\end{equation}
This representation is passed through fully connected and transposed convolutional layers to reconstruct the image $\hat{x}$:
\begin{equation}
\hat{x} = f_{\theta}(z_{\text{dec}}).
\end{equation}

\paragraph{Loss Function:}
The CVAE model is trained using a combination of a reconstruction loss and a Kullback-Leibler (KL) divergence loss. The reconstruction loss is computed as the mean squared error (MSE) between the original and reconstructed images:
\begin{equation}
\mathcal{L}_{\text{rec}} = \frac{1}{N} \sum_{i=1}^{N} \| x_i - \hat{x}_i \|^2.
\end{equation}
The KL divergence loss ensures that the learned latent distribution approximates a standard normal distribution:
\begin{equation}
\mathcal{L}_{\text{KL}} = -\frac{1}{2} \sum_{j=1}^{d} \left(1 + \log \sigma^2 - \mu^2 - \sigma^2 \right).
\end{equation}
The total loss function is given by:
\begin{equation}
\mathcal{L} = \mathcal{L}_{\text{rec}} + \lambda \mathcal{L}_{\text{KL}},
\end{equation}
where $\lambda$ is a weighting factor.

\paragraph{Training Process:}
The CVAE model is trained using the Adam optimizer with an adaptive learning rate. Stratified sampling is applied for dataset splitting to ensure balanced learning, with a $80-20$ train-validation ratio. Integer-encoded labels are used to preserve the class distribution of the images, as required by Conditional Variational Autoencoder (CVAE). The training process incorporates early stopping to prevent overfitting and learning rate scheduling to enhance convergence.

\begin{figure*}
    \centering
    \includegraphics[width=\textwidth]{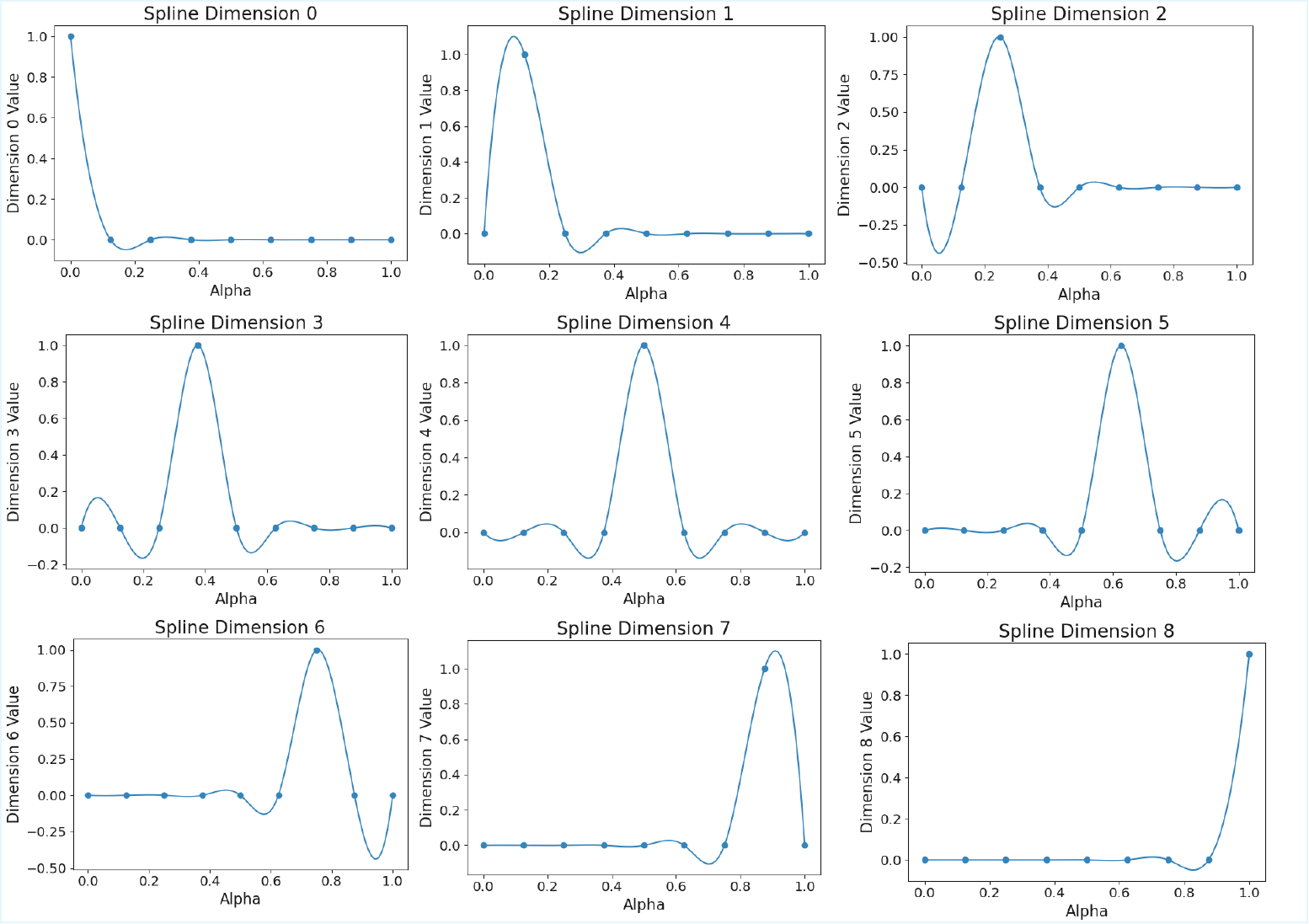}
    \caption{Visualization of cubic spline interpolation applied to the conditional label vectors in a 9-dimensional label space. Each subplot illustrates the smooth variation of one label dimension as a function of the interpolation parameter 'alpha', constructed from known one-hot encoded composition vectors corresponding to alloy compositions between 0.27 and 0.48. The peaks in each dimension reflect dominant contributions. These splines enable continuous and physically meaningful transitions between known compositions, thereby facilitating smooth label-guided image synthesis within the CVAE framework.}
    \label{fig:spline}
\end{figure*}

\section{Cubic Spline Interpolation}
\label{sec:cubic}
To generate intermediate microstructural states between known phase compositions, we employ Cubic Spline Interpolation~\cite{press2007numerical} (see Figure~\ref{fig:spline}), which ensures smooth transitions between discrete phase labels. This method is particularly useful in maintaining consistency in phase fraction variations while interpolating between microstructural states.

\paragraph{Mathematical Formulation:}
Given a set of discrete phase fraction labels:
\begin{equation}
\{ y_0, y_1, \dots, y_n \},
\end{equation}
we construct a cubic spline function \( S(x) \) that smoothly interpolates between these values. Each spline segment is defined as:
\begin{equation}
S_i(x) = a_i + b_i (x - x_i) + c_i (x - x_i)^2 + d_i (x - x_i)^3, 
 x_i \leq x \leq x_{i+1},
\end{equation}
where \( a_i, b_i, c_i, d_i \) are coefficients computed to ensure:
\begin{enumerate}[label={\roman*})]
    \item Continuity of the function: \( S_i(x_i) = y_i \).
    \item Continuity of the first derivative: \( S_i'(x_{i+1}) = S_{i+1}'(x_{i+1}) \).
    \item Continuity of the second derivative: \( S_i''(x_{i+1}) = S_{i+1}''(x_{i+1}) \).
    \item Natural boundary conditions, where the second derivative is set to zero at the endpoints.
\end{enumerate}

The dataset consists of nine discrete phase fraction values, and we define the interpolation function over normalized positions \( x \in [0,1] \):
\begin{equation}
\{ x_0, x_1, ..., x_8 \}, \quad \{ y_0, y_1, ..., y_8 \}.
\end{equation}
The spline function \( S(x) \) smoothly interpolates between these labels, generating intermediate phase fractions.

\paragraph{Selection of Labels:}
We select the labels based on target phase composition. To generate synthetic images with a specific target average phase composition \( c_{\text{avg}} \), we define a threshold tolerance:
\begin{equation}
| \text{rounded\_avg} - \text{target\_rounded\_avg} | \leq \text{tolerance}.
\end{equation}
Based on the target composition range, the corresponding interpolation interval \([x_i, x_{i+1}]\) is selected from a predefined mapping:
\begin{equation}
(x_{\text{start}}, x_{\text{end}}) = \begin{cases}
(0.27, 0.29), & 0.27 \leq c_{\text{avg}} < 0.29, \\
(0.29, 0.31), & 0.29 \leq c_{\text{avg}} < 0.31, \\
(0.31, 0.33), & 0.31 \leq c_{\text{avg}} \leq 0.33, \\
\vdots & \vdots \\
(0.47, 0.48), & 0.47 \leq c_{\text{avg}} < 0.48. \\
\end{cases}
\end{equation}
A set of interpolated labels is then sampled from \( S(x) \) within this range.

\paragraph{Image Generation:}
Once the interpolated labels are obtained, we generate images by sampling a Gaussian latent vector:
\begin{equation}
z \sim \mathcal{N}(0, I).
\end{equation}
The interpolated label is converted into a tensor and passed into the CVAE decoder, which reconstructs the corresponding microstructure:
\begin{equation}
\hat{x} = h_{\phi}(z, \tilde{y}),
\end{equation}
where \( \tilde{y} \) is the interpolated phase label.

To ensure the generated image has the desired c\textsubscript{avg}, we compute the normalized average pixel intensity:
\begin{equation}
\bar{c} = \frac{1}{HW} \sum_{i=1}^{H} \sum_{j=1}^{W} x_{ij},
\end{equation}
where \( H \) and \( W \) are the image dimensions. If \( |\bar{c} - c_{\text{avg}}| \leq \text{tolerance} \), the image is accepted; otherwise, the process is repeated.

\paragraph{Morphological Characterization:}
We perform further image analysis to assess the microstructural variations in the dataset generated via cubic spline interpolation. This analysis aims to arrange images in terms of the size of the microstructural features. First, each generated (from cubic spline interpolation) microstructure image is converted to a binary format using thresholding:
\begin{equation}
B(I) = 
\begin{cases} 
255, & \text{if } I(x,y) \geq 127, \\
0, & \text{otherwise}.
\end{cases}
\end{equation}
Contours corresponding to distinct microstructural regions are extracted using OpenCV’s contour detection algorithm. Given \( N_c \) detected contours in an image, the total shape area is computed as:
\begin{equation}
A_{\text{total}} = \sum_{i=1}^{N_c} A_i,
\label{eqatot}
\end{equation}
where \( A_i \) represents the area of the \( i \)-th detected microstructural shape. Finally, the average feature (denoted by white shapes) size per image is calculated as:
\begin{equation}
A_{\text{avg}} = 
\begin{cases} 
\frac{A_{\text{total}}}{N_c}, & N_c > 0, \\
0, & \text{if } N_c = 0.
\end{cases}
\label{eqavg}
\end{equation}
The average feature size for a given composition increases with time because of coarsening. We rank the images (generated from cubic spline interpolation) based on their computed feature size \( A_{\text{avg}} \). Three key images are identified:
\begin{enumerate}[label=\roman*)]
\item Smallest average feature size: \(I_{\text{min}}=\arg\min A_{\text{avg}}\).
\item Largest average feature size: \(I_{\text{max}}=\arg\max A_{\text{avg}}\).
\item Medium average feature size: \(I_{\text{mid}}\), corresponding to the median rank in the sorted dataset.
\end{enumerate}
These three images are further used to get the time evolution of microstructures. 

Note that while cubic spline interpolation generates microstructure images for a targeted composition value, it does not ensure smooth morphological transitions with time. For a given target composition, between the latent vectors of images with the smallest and largest feature size, as obtained from the cubic interpolation, SLERP (Spherical Linear Interpolation) is applied to generate intermediate steps. This method ensures smooth morphological transitions with time, very similar to the phase-field time evolution. Thus, both the interpolation methods are necessary and complementary. Further details about SLERP are given in the following section.

\begin{figure}
    \centering
    \includegraphics[width=0.4\textwidth]{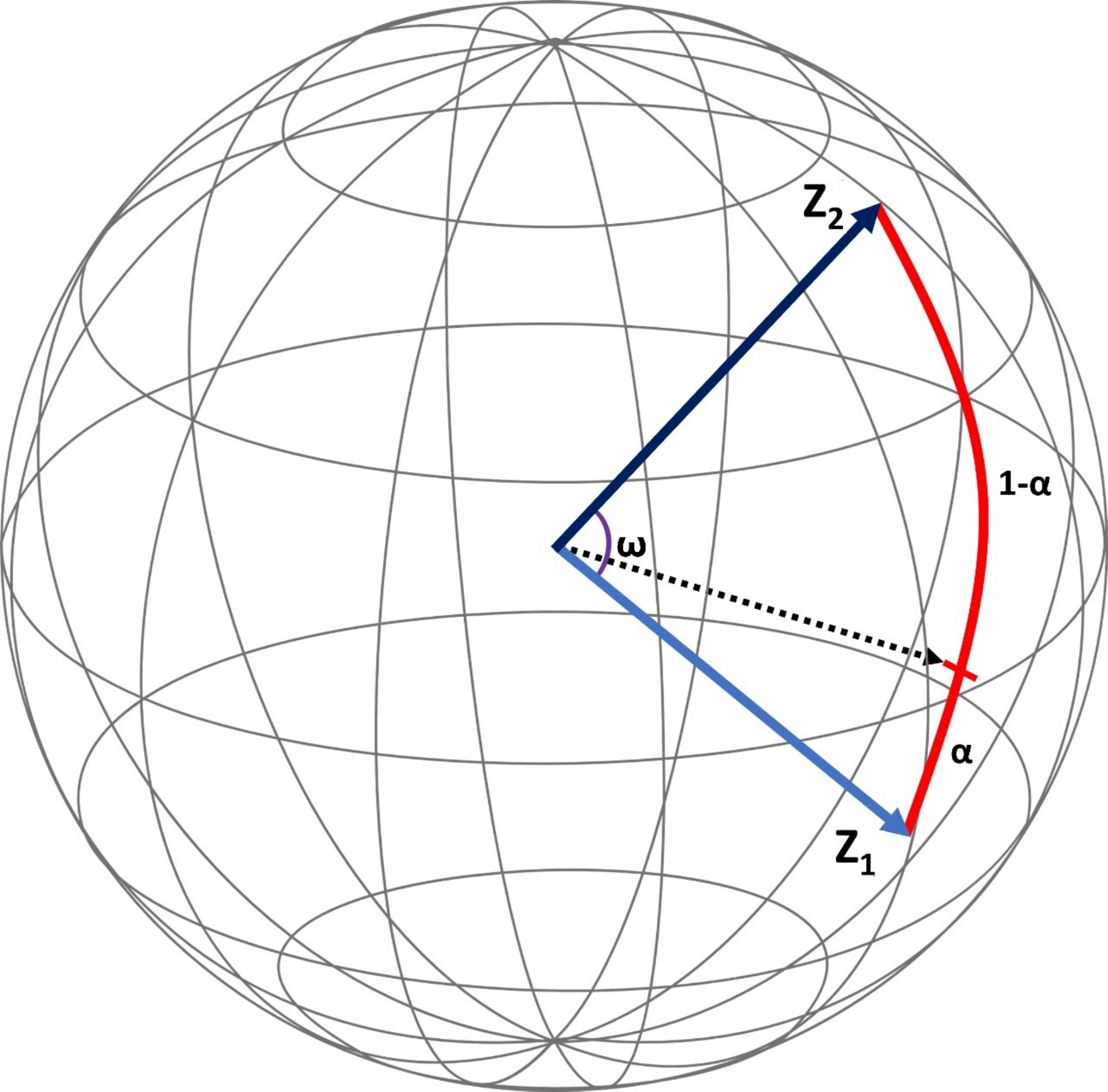}
    \caption{Illustration of Spherical Linear Interpolation (SLERP). Given two latent vectors \( z_1 \) and \( z_2 \), SLERP computes an intermediate point along the geodesic path on the unit hypersphere. The interpolation factor \( \alpha \) determines the position of the generated latent vector, ensuring smooth and consistent transitions in the latent space.}

    \label{fig:slerp}
\end{figure}

\section{Spherical Linear Interpolation (SLERP) for Microstructure Evolution}
\label{sec:slerp}
To capture the smooth evolution of microstructures between different morphological states, we employ Spherical Linear Interpolation (SLERP)~\cite{shoemake10.1145/325334.325242} in the latent space of the Conditional Variational Autoencoder (CVAE). This method (see Figure~\ref{fig:slerp}) ensures a geodesic transition between latent representations, preventing distortions caused by linear interpolation in high-dimensional spaces. 

\paragraph{Mathematical Formulation:}
Given two latent vectors \( z_1 \) and \( z_2 \) in a high-dimensional space, the Spherical Linear Interpolation (SLERP) function computes an intermediate latent representation \( z_{\text{interpolated}} \) at a specific interpolation factor \( \alpha \):
\begin{equation}
z_{\text{interpolated}}(\alpha) = \frac{\sin((1 - \alpha) \omega)}{\sin \omega} z_1 + \frac{\sin(\alpha \omega)}{\sin \omega} z_2,
\end{equation}
where the angular distance between the two latent vectors is:
\begin{equation}
\omega = \arccos \left( \frac{z_1 \cdot z_2}{\|z_1\| \|z_2\|} \right).
\end{equation}
The interpolation parameter \( \alpha \) varies in the range \( \alpha \in [0,1] \), where:
\begin{enumerate}[label=\roman*)]
    \item \(\alpha = 0\): returns the initial latent vector \(z_1\).
    \item \(\alpha = 1\): returns the final   latent vector \(z_2\).
    \item \(0<\alpha<1\): yields smooth intermediate latent vectors.
\end{enumerate}

\begin{figure*}
        \centering

    \subfigure[\( c_{\text{avg}} = 0.27 \rightarrow 0.39 \)]{
        \centering
        \includegraphics[width=\textwidth]
                {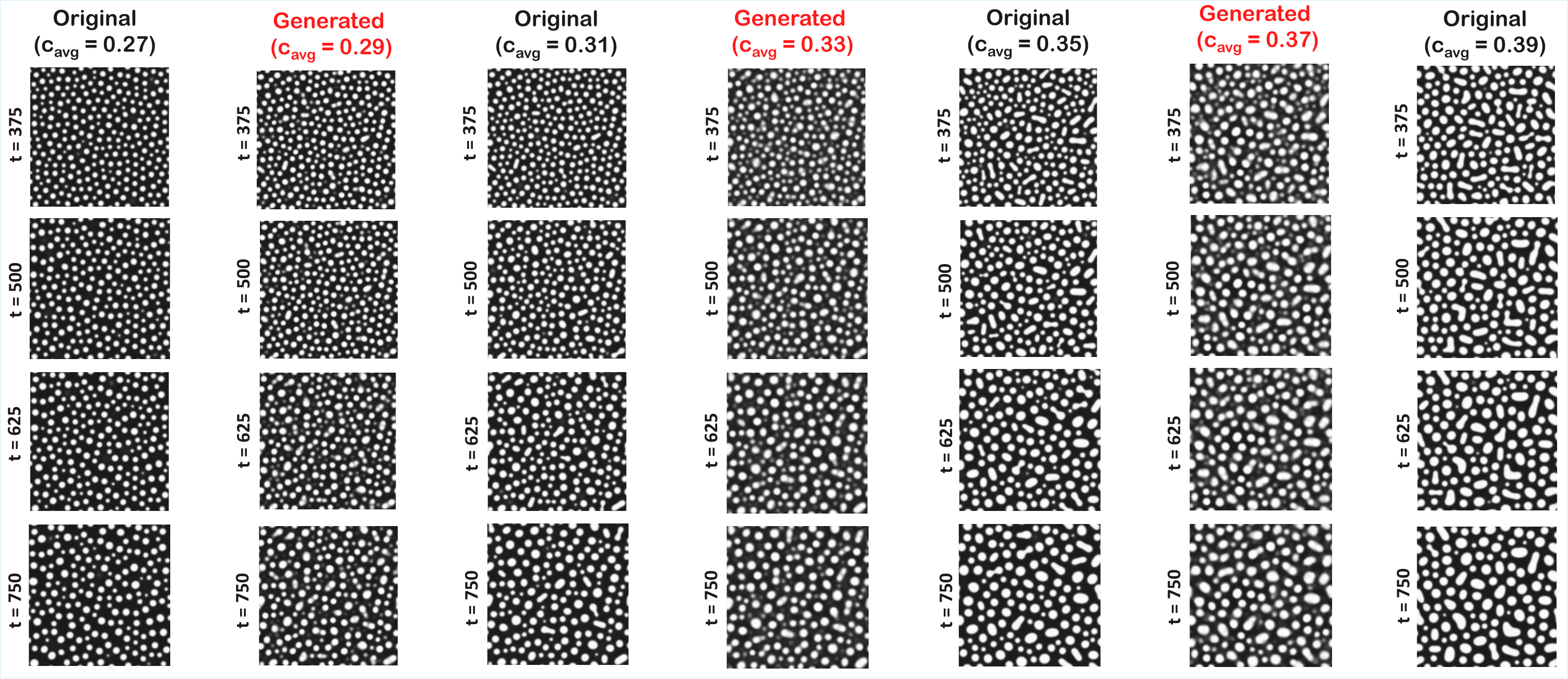}
        \label{fig:comparison_a}}
    \hfill
    \subfigure[\( c_{\text{avg}} = 0.40 \rightarrow 0.46 \)]{
        \centering
        \includegraphics[width=\textwidth]
                {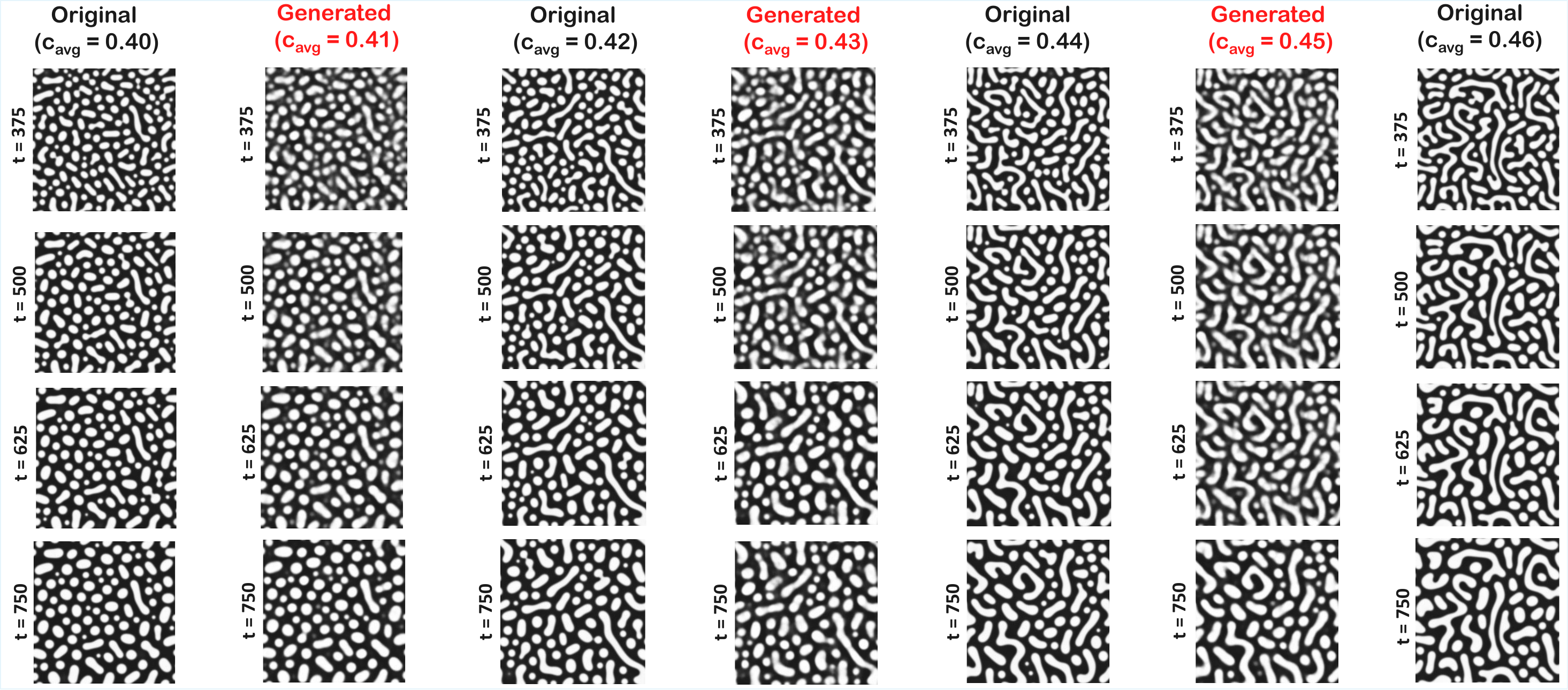}
        \label{fig:comparison_b}}
\hfill
    \caption{Comparison of original and CVAE-generated microstructure
             evolution for two composition intervals: (a)  
             \(0.27 \rightarrow 0.39\) and (b) \(0.40 \rightarrow 0.46\).}
    \label{fig:comparison}
\end{figure*}

\begin{figure*}
    \centering
    \includegraphics[width=\textwidth]{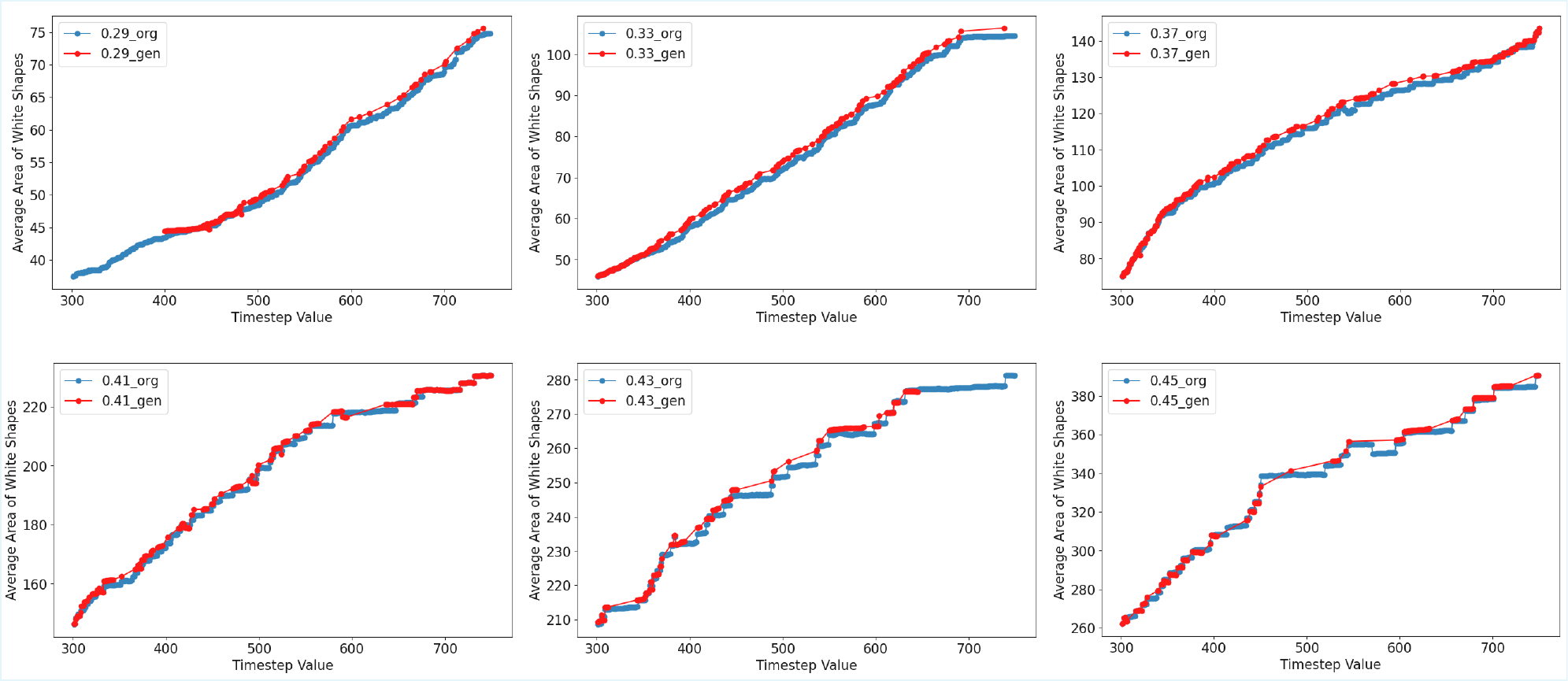}
    \caption{Timestep vs. average area plot of microstructure evolution for original and generated microstructures across varying average compositions. Each subplot corresponds to a distinct composition: the blue curve represents the original microstructure, and the red curve denotes the generated counterpart.}
    \label{fig:autocorr1}
\end{figure*}

\paragraph{Image Generation:}
As mentioned in Section~\ref{sec:cubic}, we first identify three key images, $I_{\text{min}}, I_{\text{mid}}, I_{\text{max}}.$ Next, we perform SLERP-based latent space interpolation in two stages:
\begin{enumerate}[label=\roman*)]
\item Between \( I_{\text{min}} \) and \( I_{\text{mid}} \) to generate intermediate images covering the transition from small to medium feature size.
\item Between \( I_{\text{mid}} \) and \( I_{\text{max}} \) to generate intermediate images covering the transition from medium to large feature size.
\end{enumerate}
The above method ensures that the entire transition from small to large feature size is captured smoothly for a given target composition.

To obtain latent representations, the CVAE encoder extracts the mean and variance of the Gaussian distribution for the terminal images. For example, in the case of $I_{\text{min}}$ and $I_{\text{mid}}$,
\begin{eqnarray}
\mu_1, \log \sigma_1^2 = f_{\theta}(I_{\text{min}}, y_1),\\\nonumber
\mu_2, \log \sigma_2^2 = f_{\theta}(I_{\text{mid}}, y_2).
\end{eqnarray}
Latent vectors \( z_1 \) and \( z_2 \) are sampled using the reparameterization trick:
\begin{eqnarray}
z_1 = \mu_1 + \exp\left(\frac{\log \sigma_1^2}{2}\right) \cdot \epsilon,\\\nonumber
z_2 = \mu_2 + \exp\left(\frac{\log \sigma_2^2}{2}\right) \cdot \epsilon.
\end{eqnarray}
Interpolated latent representations are computed using SLERP at multiple values of \( \alpha \):
\begin{equation}
z_{\text{interpolated}} = \text{SLERP}(\alpha, z_1, z_2),
\end{equation}
where \( \alpha \) is sampled from a uniform distribution between 0 and 1 to generate a sequence of smooth transitions.

Finally, each interpolated latent representation \( z_{\text{interpolated}} \) is passed through the CVAE decoder to reconstruct microstructures:
\begin{equation}
\hat{x} = h_{\phi}(z_{\text{interpolated}}, y_{\text{interpolated}}),
\end{equation}
where \( y_{\text{interpolated}} \) is the interpolated phase label, computed as:
\begin{equation}
y_{\text{interpolated}} = (1 - \alpha) y_1 + \alpha y_2.
\end{equation}

\paragraph{Morphological Characterization:}
To ensure physical consistency, we evaluate the generated images based on their average feature area \( A_{\text{avg}} \), as described in Eq.~\ref{eqatot} and Eq.~\ref{eqavg}.
The interpolation process is accepted provided the following conditions are satisfied:
\begin{enumerate}[label=\roman*)]
\item Absolute difference between the average area of the first generated image and I$_{min}$ is less than 2, i.e., $| A_{\text{gen,first}} - A_{\text{min}} | < 2.$
\item Absolute difference between the average area of the last generated image and I$_{max}$ is less than 2, i.e., $| A_{\text{gen,last}} - A_{\text{max}} | < 2.$
\end{enumerate}
If any one of the conditions is not met, the present set of generated microstructures is discarded and SLERP is rerun to generate another set of images. This process ensures morphological consistency in the microstructure evolution process. Model performance is analyzed in detail in the following section.


\begin{figure*}
    \centering
    \includegraphics[width=\textwidth]{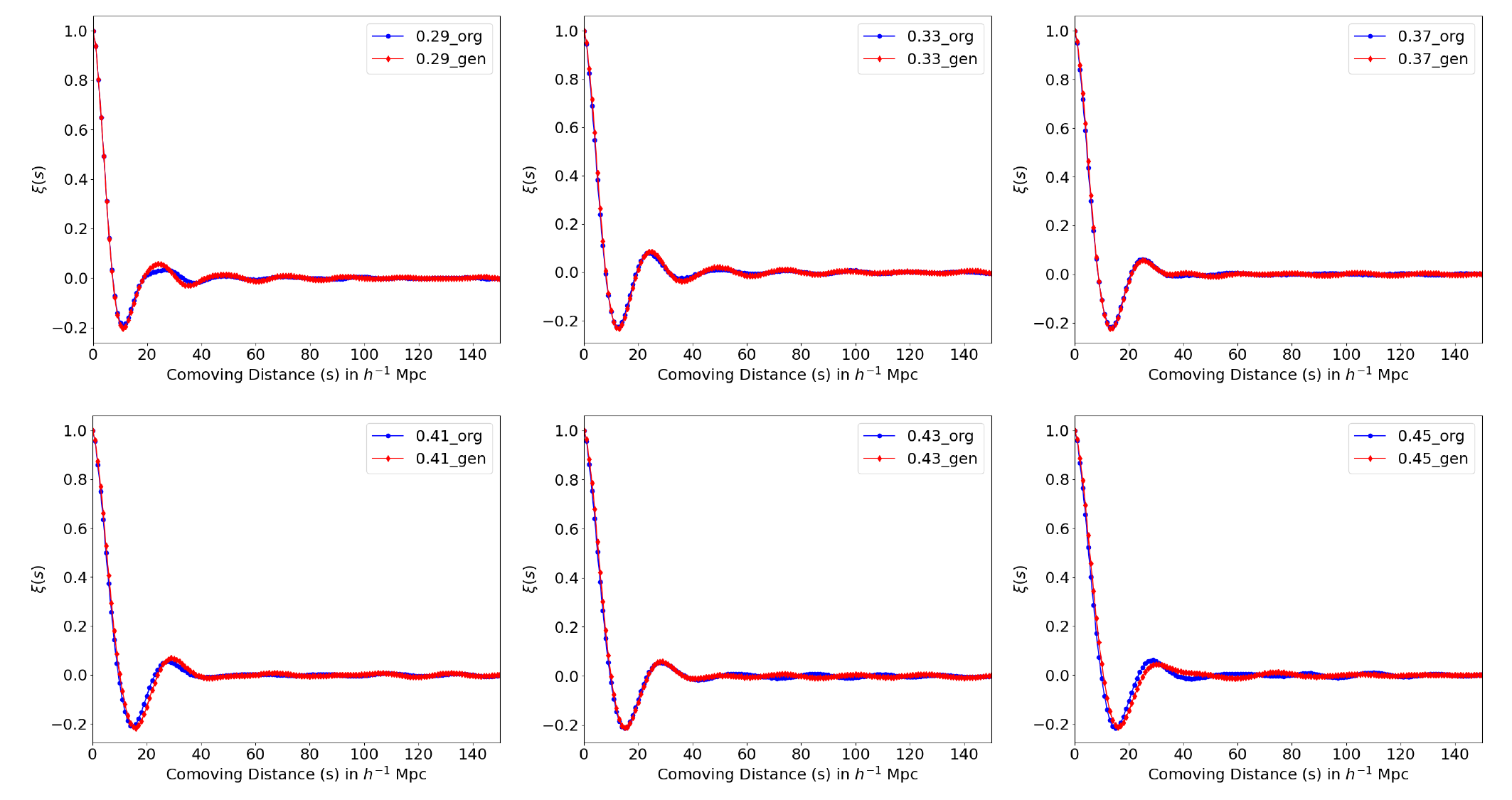}
    \caption{2-point autocorrelation compared at the 750$^{th}$ timestep between original and generated microstructures. Each subplot corresponds to a distinct composition, where the blue (red) curve represents the 2-point autocorrelation for the original phase-field (CVAE-generated) microstructure. The high degree of overlap across all subplots indicates that the generative model successfully captures the spatial correlation characteristics of the original data.}
    \label{fig:autocorr2}
\end{figure*}

\section{Results and Discussions}
\label{sec:discussions}
So far, we have shown a step-by-step development of a Conditional Variational Autoencoder (CVAE) model for generating microstructure images conditioned on varying compositions. Cubic spline and spherical linear interpolation in the latent space generate the composition and time variation. The method predicts microstructure evolution for intermediate targeted composition values, based on a training dataset comprising a few compositions. This results in a significant computational speedup compared to the case of spanning the entire composition range via the phase-field approach. It takes $\approx 60$ minutes to train the model on an Intel\textsuperscript{\textregistered} Xeon\textsuperscript{\textregistered} Gold 6338N CPU (2.20~GHz), supported by 128 GB of RAM. The model is faster than conventional numerical techniques for solving the Cahn-Hilliard equation. While the traditional numerical technique for phase-field calculation takes $\approx 12$ minutes to generate 1000 frames of the microstructure evolution for a given composition, the ML-assisted approach requires only $\approx 3$ minutes for the same.

Figure~\ref{fig:comparison} and Figures~S1-S3 (Supplementary Material) compare the original training compositions and the generated intermediate target compositions obtained by deep learning. Clearly, the method can produce high-resolution images showing smooth time evolution of microstructures, very similar to actual phase-field simulations. After visually confirming our model's effectiveness in generating phase-field-like microstructure evolution, let us perform a more quantitative analysis across two key metrics: the temporal evolution of feature size and spatial correlation similarity.

Figure~\ref{fig:autocorr1} compares the CVAE-generated microstructure evolution and the phase-field model evolution across different compositions. The horizontal axis represents timestep values, while the vertical axis shows the average area of the features (white in color) in the microstructure at each timestep. The area increases with time because of coarsening. The blue curves represent the microstructure evolution via the phase-field model, and the red curves represent the same via CVAE-based deep learning. We observe that for all compositions tested, both models exhibit similar trends of coarsening over time. The close alignment between the two curves indicates that the CVAE model can learn and replicate the temporal dynamics of microstructure evolution like traditional phase-field simulations. This is an important finding, as it shows that the model can generate physically plausible microstructure evolution, very similar to the actual phase-field simulations. Additionally, the model's ability to create microstructures with varying particle sizes across different compositions highlights its potential for use in material design and alloy optimization, where controlling particle size and distribution is critical.

Figure~\ref{fig:autocorr2} compares the 2-point autocorrelation function~\cite{torquato2002random} between the phase-field and CVAE-generated microstructures. The autocorrelation function measures the spatial similarity between microstructure features at varying distances, which indicates grain distribution and phase homogeneity in the material. In the figure, the blue lines represent the autocorrelation functions for the phase-field model, and the red lines represent the autocorrelation functions for the CVAE-generated microstructures. The spatial decay patterns are similar for each composition between the two models. This suggests that the CVAE-generated microstructures not only follow the same evolutionary trends over time but also exhibit comparable spatial structures to the phase-field microstructures. While the CVAE model exhibits minor differences in autocorrelation decay, particularly at larger distances, the overall similarity confirms that the model can generate realistic spatial patterns of grains and phases. These findings further support the utility of the CVAE model for generating realistic microstructures for material science applications, where understanding spatial correlations is key to predicting material properties.

\section{Conclusions}
\label{sec:conclusions}
This study demonstrates the application of a Conditional Variational Autoencoder (CVAE) for generating microstructure images conditioned on varying alloy compositions. The use of latent space interpolation techniques (Cubic Spline and SLERP) enables the model to generate intermediate microstructures across varying compositions, making the model very useful in exploring a large number of alloy compositions in a short time. A detailed comparison between phase-field and deep-learned microstructures proves that the CVAE model can replicate the overall characteristics of microstructure evolution, including both temporal and spatial features. Key findings from this work include:
\begin{enumerate}[label=\alph*)]
    \item The Timestep vs Average Area plot (Figure~\ref{fig:autocorr1}) confirms that the CVAE model effectively generates microstructures that evolve similarly to phase-field simulations, with corresponding grain coarsening patterns across timesteps.
    \item The 2-Point Autocorrelation Comparison (Figure~\ref{fig:autocorr2}) demonstrates that the CVAE model generates spatially realistic microstructures similar to phase-field simulations, highlighting the deep-learning model's potential for capturing spatial distribution patterns of particles.
    \item The model's ability to generate physically plausible microstructures opens up new possibilities for material design and alloy optimization, where understanding and controlling microstructure at different stages of evolution is critical.
    \item Overall, the results affirm the effectiveness of the CVAE model in generating realistic microstructures with temporal and spatial features comparable to phase-field simulations, making it a promising tool for materials informatics and design optimization.
\end{enumerate}

    
    


\section{Acknowledgement}
We acknowledge the National Super Computing Mission (NSM) for providing computing resources of ``PARAM Sanganak'' at IIT Kanpur, which is implemented by CDAC and supported by the Ministry of Electronics and Information Technology (MeitY) and Department of Science and Technology (DST), Government of India. We also thank the ICME National Hub, IIT Kanpur, and CC, IIT Kanpur, for providing the HPC facility.

\bibliographystyle{model1-num-names}

\end{document}